\documentclass{appolb}
\usepackage{graphicx}
\usepackage{epsf,amsmath,amssymb,graphicx,dcolumn,url}
\usepackage{subcaption}
\usepackage{scalefnt}

\def\bal#1\eal{\begin{align}#1\end{align}}

\newcommand{\abbrev}{\scalefont{.9}}

\newcommand{\mh}{m_{h}}

\newcommand{\mtop}{m_{\rm top}}

\newcommand{\lhc}{{\abbrev LHC}}

\newcommand{\powheg}{{\abbrev POWHEG}}

\newcommand{\pt}{\ensuremath{p_T}}
\newcommand{\ptjet}{\ensuremath{p_T^{\text{jet}}}}

\newcommand{\ptmin}{\ensuremath{p_{T,\text{min}}^{\text{jet}}}}

\newcommand{\nll}{{\abbrev NLL}}
\newcommand{\nnll}{{\abbrev NNLL}}
\newcommand{\lo}{{\abbrev LO}}
\newcommand{\nlo}{{\abbrev NLO}}
\newcommand{\nnlo}{{\abbrev NNLO}}
\newcommand{\nnnlo}{{\abbrev N3LO}}
\newcommand{\sm}{{\abbrev SM}}

\newcommand{\ps}{{\abbrev PS}}

\newcommand{\mb}{m_{\rm b}}

\newcommand{\plus}{{\abbrev +}}
\newcommand{\citere}[1]{Ref.\,\cite{#1}}
\newcommand{\citeres}[1]{Refs.\,\cite{#1}}
\newcommand{\eqn}[1]{Eq.\,(\ref{#1})}

\newcommand{\fig}[1]{Fig.\,\ref{#1}}

\newcommand{\dd}{{\rm d}}

\newcommand{\als}{\ensuremath{\alpha_s}}

\newcommand{\muF}{\mu_{\rm F}}
\newcommand{\muR}{\mu_{\rm R}}

\newcommand{\pdf}{{\abbrev PDF}}

\newcommand{\mcatnlo}{{\abbrev MC@NLO}}

\begin{document}

\title{A brief theory overview of Higgs physics at the LHC%
\thanks{Presented at the {\abbrev XXXIX} International Conference of Theoretical Physics ``Matter to the Deepest", Ustroń, Poland, September 13--18, 2015.}%
}
\author{Marius Wiesemann
\address{Physik-Institut, Universit\"at Z\"urich, CH-8057 Z\"urich, Switzerland}
}
\maketitle
\begin{abstract}
This is a brief review of the theoretical status of Higgs production 
at the \lhc{} in
the Standard Model, with an emphasis on the recent 
developments and computations. In particular, I focus on both inclusive and 
differential cross sections for the dominant production 
mode in the Standard Model, 
where  the theoretical uncertainties induced by the different 
interplays between top- and 
bottom-quark effects in the gluon-fusion scattering amplitude are discussed. 
\end{abstract}
  
\section{Introduction}
The discovery of a scalar particle \cite{Aad:2012tfa,Chatrchyan:2012xdj} 
is already considered as the legacy of the 
\lhc{} Run I. Several studies have analyzed the properties of this resonance 
(see, e.g., Refs.\,\cite{Aad:2015mxa,Khachatryan:2014jba}), all of which are fully consistent with the Standard Model 
(\sm{}) picture.
Such studies rely heavily on the precise theoretical understanding of the Higgs 
production and decay mechanisms.

Already with Run I, but even more 
with the start of Run II, the \lhc{} is entering precision physics, in particular for 
Higgs related observables. This accuracy enables the measurement of differential 
distributions of the Higgs boson. First differential results were recently published in 
Refs.\,\cite{Aad:2015lha,Khachatryan:2015rxa}. With the additional precision expected from Run II such measurements 
will allow for the determination of small deviations from the \sm{} picture in order to 
test the nature of the Higgs couplings. Therefore, accurate theoretical predictions
are required not only for the total rate, but also for differential cross sections.

This contribution reviews the state-of-the-art predictions in the \sm{} for the dominant 
gluon-induced Higgs production mode, where the total inclusive 
cross section 
as well as fully-exclusive observables will be discussed.

\section{Higgs production through gluon fusion}
\label{sec:gf}

Standard Model Higgs production proceeds predominantly via
gluon fusion, where the Higgs-gluon coupling is mediated by a quark
loop. 
Its cross section is roughly one order of magnitude larger than the
sum of all other production modes. In the past years, gluon-induced Higgs 
production has become the theoretically best studied process at 
hadron colliders, which lead to a significant decrease of the related theoretical
uncertainties. 

An effective field theory approach, where the top quark is assumed to be 
infinitely heavy (also known as heavy-top limit), allows to 
determine higher order corrections roughly one perturbative order higher than in the 
full theory.
One must bear in mind, however, that this approximation is strictly 
valid only when all scales remain below twice the top-quark mass. Therefore, care 
must be taken for the total inclusive cross section by estimating the effect of missing 
quark-mass contributions and 
for kinematical distributions outside the validity range of the 
approximation. In the latter case, the full theory must be employed at the cost 
of lower perturbative accuracy.

\subsection{Total inclusive cross section}

For a long time, the highest perturbative accuracy available 
for the total inclusive Higgs cross section in the 
infinite-top mass approximation  was next-to-next-to-leading order (\nnlo{})
\cite{Harlander:2002wh,Anastasiou:2002yz,Ravindran:2003um}. 
Very recently, a milestone in perturbative computations was reached, when 
the first next-to-\nnlo{} (\nnnlo{}) result at hadron colliders was computed 
\cite{Anastasiou:2015ema}
for this process. Similar to the \nnlo{} result 
of \citere{Harlander:2002wh} the computation of 
\citere{Anastasiou:2015ema} employs a
threshold expansion (including the first 39 terms) 
in $\tau=\mh^2/s$ around $\tau=1$, where $\mh$ is the 
mass of the Higgs boson and $s$ the partonic center-of-mass energy.
While the leading terms of this expansion
\cite{Anastasiou:2014vaa,Anastasiou:2014lda}
do actually not yet capture the dominant contribution of the \nnnlo{} corrections, 
the expansion stabilizes after $\sim 10$ terms \cite{Anastasiou:2015ema}. 
This is indeed very similar to what was observed at \nnlo{} \cite{Harlander:2002wh}, 
which later turned out to be in excellent agreement with the full result. 
There are essentially two reasons why the expansion works so well: 
soft-gluon effects close to threshold are important, which are captured by this 
approach, and the gluon luminosities strongly suppress the region $z\ll 1$, where 
the expansion looses its validity. Hence, the computation 
of \citere{Anastasiou:2015ema}
can be safely considered to be the full \nnnlo{} result for all practical purposes.

The size of the \nnnlo{} corrections remains rather small ($\lesssim 1\%)$, once 
a judicious choice for the renormalization and factorization scales of 
$m_h/2$ at \nnlo{} is made.
This scale choice was motivated already by the well 
agreement with the soft-gluon resummed result 
at \nnlo{}\plus{}\nnll{} \cite{deFlorian:2009hc}.
The actual benefit of the \nnnlo{} corrections is a significant 
reduction of the residual uncertainties leading to a prediction with a 
precision at the level of $2-3$\%.

Consequently, at \nnnlo{} the result is perturbatively highly stable, 
while all other uncertainties require now a careful estimate:
\pdf{} uncertainties are already of similar size at next-to-leading order (\nlo{}) 
and at \nnlo{}, which 
should not change at \nnnlo{}, since no dedicated \nnnlo{} \pdf{} sets exist. They 
amount to typically $\sim 5\%$ and have therefore become one of the major 
uncertainties on the cross section prediction for hadronic Higgs production
at \nnnlo{}.
Electro-weak effects have been evaluated for example in
\citeres{Degrassi:2004mx,Aglietti:2004nj,Actis:2008ug,
Anastasiou:2008tj}. The uncertainty induced by neglecting finite top-mass 
effects on the radiative corrections in the heavy-top approximation 
has been estimated to be below 
1\% for the total cross sections at \nnlo{} by studying the asymptotic expansion 
in inverse powers of the top-quark mass ($1/m_{\rm top}^2$) 
\cite{Harlander:2009my,Pak:2011hs}. One should bear 
in mind, however, that top mass effects must be fully accounted for at leading order 
and can then be supplemented by radiative corrections computed in the heavy-top 
approximation in terms of $K$-factors.
For the partonic cross section the $1/m_{\rm top}^2$ expansion reads
\bal
\hat\sigma= \sum\limits_{k=0}^\infty \,\frac1{\mtop^{2k}}\, \hat\sigma^	{(k)},
\eal
which is strictly valid only for $\sqrt{s}<2\,\mtop$. Since the coefficients of 
this expansion actually diverge as $\sqrt{s}\rightarrow \infty$
\cite{Harlander:2009my},
a matching of the partonic cross section to the high-energy limit 
$\hat\sigma(\sqrt{s}\rightarrow\infty)$ \cite{Marzani:2008az,Harlander:2009my} 
was introduced in 
\citere{Harlander:2009my}.\footnote{
Note that the region $\sqrt{s}>2\,\mtop$ is strongly suppressed by the gluon 
luminosities.}
The dependence of the estimated top-mass effects on the precise details of the matching procedure is negligible.

Concerning other quarks as mediators of the gluon-Higgs coupling, the
effect of the four lightest quarks is $\lesssim 1$\%.
The bottom quark, on the other hand,
contributes $\sim5-10$\% to the total cross section at \nlo{}
\cite{Spira:1995rr}. 
Due to the small value of the bottom mass, a heavy-quark approximation 
as used for the top-quark contributions is not suitable in this case.
Therefore, bottom-quark effects must be included solely 
at the perturbative order, where the full quark-mass dependence on the cross 
section is known. Nevertheless, the uncertainty induced by
the missing bottom-quark contributions at \nnnlo{} can simply be estimated 
by assuming their $K$-factor
to be not larger than the one used for the top-quark contribution.\footnote{Radiative corrections to the bottom loop are assumed to be smaller 
than for the top loop due to the softer spectrum. At \nlo{} this already turned 
out to be true.} This leads to an estimate of at most $\pm 2-3$\% missing 
bottom-quark effects at \nnnlo{}.

In conclusion, the theoretical prediction for the total inclusive cross section for 
Higgs production through gluon fusion is under excellent theoretical control, which 
allows for actual precision physics at \lhc{} Run II.

\subsection{Differential cross sections at fixed order}

Kinematical distributions in hadronic Higgs production provide an important 
handle on the determination of Higgs properties. 
Among the most relevant observables in this respect is the
Higgs transverse momentum ($\pt$) spectrum.
The \nlo{} transverse momentum distribution of the Higgs boson in
gluon fusion at $\pt{}>0$ has been known for some
time in the limit of heavy top quarks \cite{deFlorian:1999zd,Glosser:2002gm}. 
Recently, \nnlo{} corrections to this observable 
were determined in a fully-differential 
computation of the Higgs+jet rate 
\cite{Boughezal:2013uia,Chen:2014gva,Boughezal:2015dra,
Boughezal:2015aha,Caola:2015wna}. Perturbative effects on 
the \pt{} spectrum at \nnlo{} turn out to 
be quite sizable; they amount to $\sim 20\%$ in the tail of the distribution.

\begin{figure}[t]
\begin{center}
\includegraphics[width=0.7\textwidth]{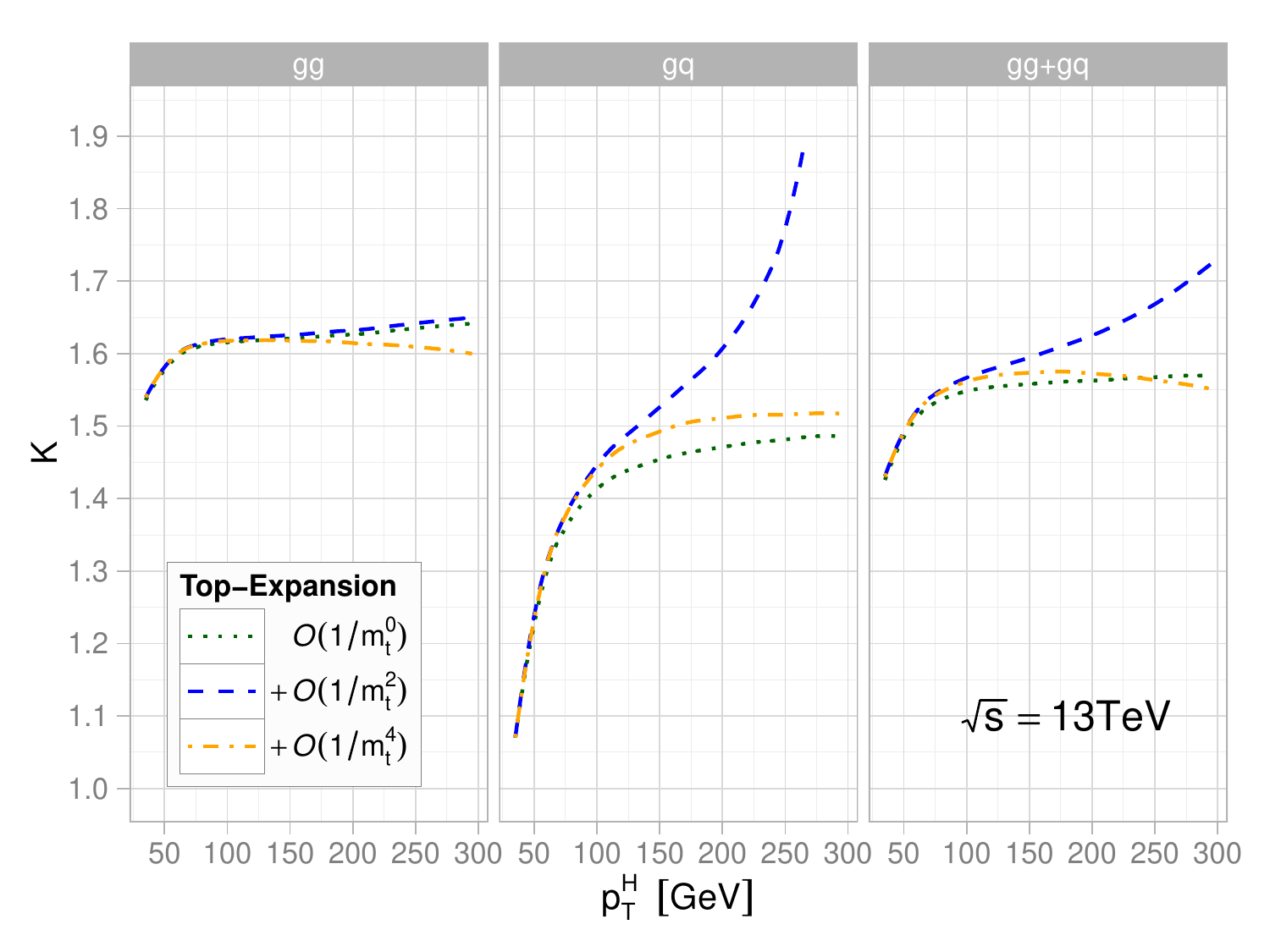}
    \parbox{\textwidth}{
      \caption[]{\label{fig:pTH}{$K$-factors for the Higgs \pt{}
       distribution as defined in \eqn{eq:K}, i.\,e. $K \equiv
    K_k^{\rm\nlo{}}(\pt)$. Left/center/right plot: only $gg$/only $qg$/sum of $gg$
    and $qg$. Dotted/dashed/dash-dotted: $k = 0/2/4$. } }}
\end{center}
\end{figure}

A careful assessment of the validity of the heavy-top approximation becomes 
even more important for differential observables, where the hierarchy of different 
scales is more involved. Subleading top-mass effects on the differential 
Higgs+jet cross section at \nlo{} have been considered in
\citeres{Harlander:2012hf,Neumann:2014nha}. \fig{fig:pTH} shows the differential
$K$-factor for the Higgs \pt{} distribution\footnote{$[\ldots]_{\mtop^k}$ denotes 
the truncation of the the asymptotic expansion at $1/\mtop^k$.}
\bal
\label{eq:K}
K_k^{\rm\nlo{}}(p_T)\equiv [\dd\sigma^{\rm \nlo{}}/\dd\pt{}]_{\mtop^k}\Bigm/[\dd\sigma^{\rm \lo{}}/\dd\pt{}]_{\mtop^k}
\eal
as an expansion up to $1/\mtop^k$, 
where $k=0$ (green, 
dotted curve) corresponds to the heavy-top limit, $k=2$ (blue, dashed curve) 
involves terms up to $1/\mtop^2$ and $k=4$ (yellow, dash-dotted curve)
 includes the $1/\mtop^4$ term in addition. 
The factorization and renormalization scales are set to the transverse mass of the Higgs boson $\muF^2=\muR^2=m_H^2+\pt^2$.
  The left panel 
of \fig{fig:pTH} shows the purely gluon-induced subchannel, where the convergence 
of the asymptotic expansion is close to excellent at least for $\pt\lesssim 150$\,GeV.
This consistent picture deteriorates to some extend for the gluon-quark induced 
channel (central panel), but leads in the sum of both channels (right panel) to an 
overall reasonable convergence of the asymptotic series as long as 
$\pt\lesssim \mtop$. In that region the uncertainty associated with missing 
top-mass effects in the heavy-top limit can be estimated to be below $2-3\%$, when 
taking the spread of the different curves as a measure of the size of the top-quark 
effects. At $\pt= 300$\,GeV the uncertainty is already at the level of $10$\% and 
the heavy-top limit starts to become unreliable.

\begin{figure}[t]
	\begin{subfigure}[b]{0.5\textwidth}
         	\includegraphics[width=\textwidth]{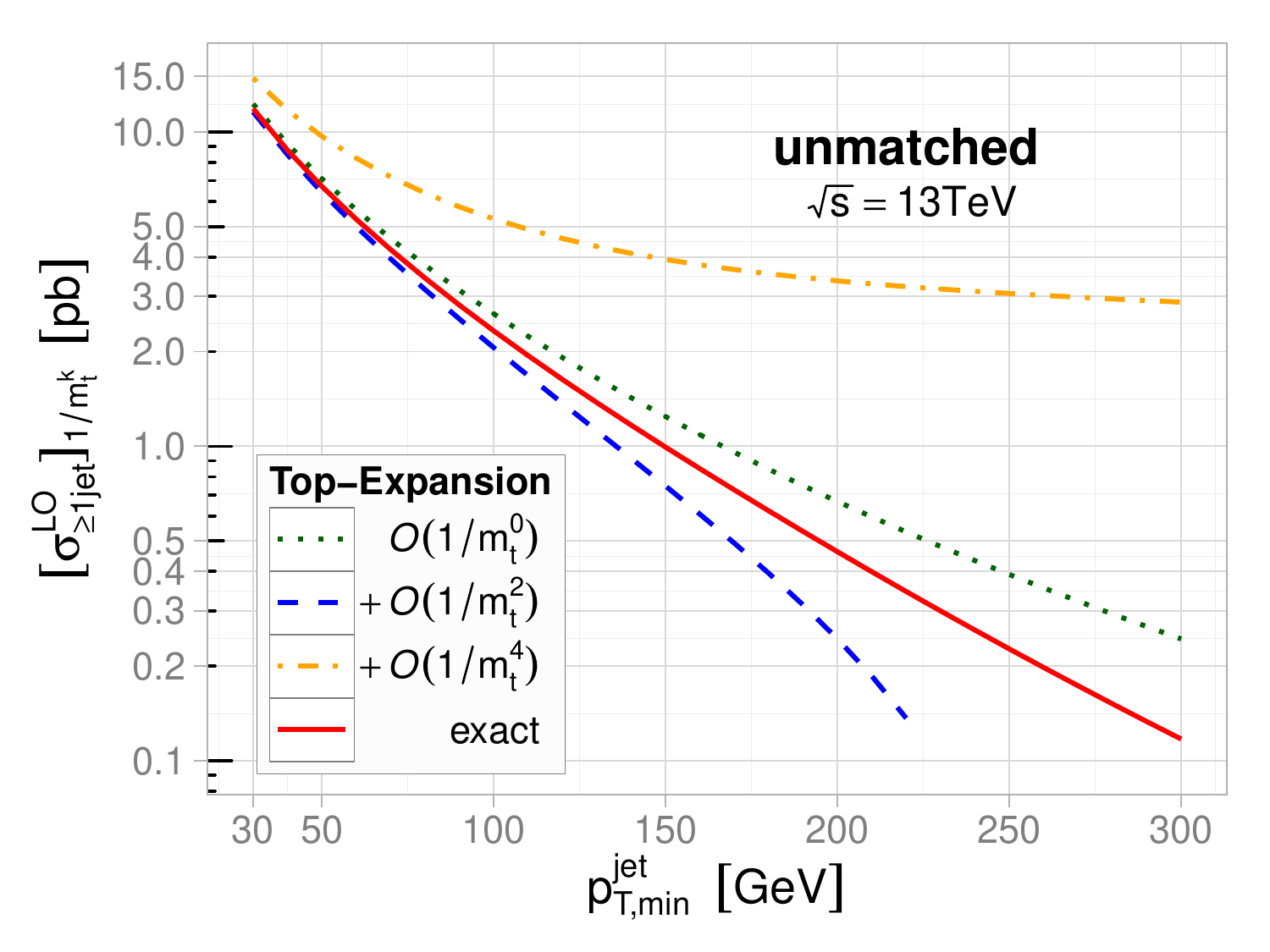}
	        \caption{}
            \label{fig:onejetLO_unmatched}
        \end{subfigure}%
	\begin{subfigure}[b]{0.5\textwidth}
         	\includegraphics[width=\textwidth]{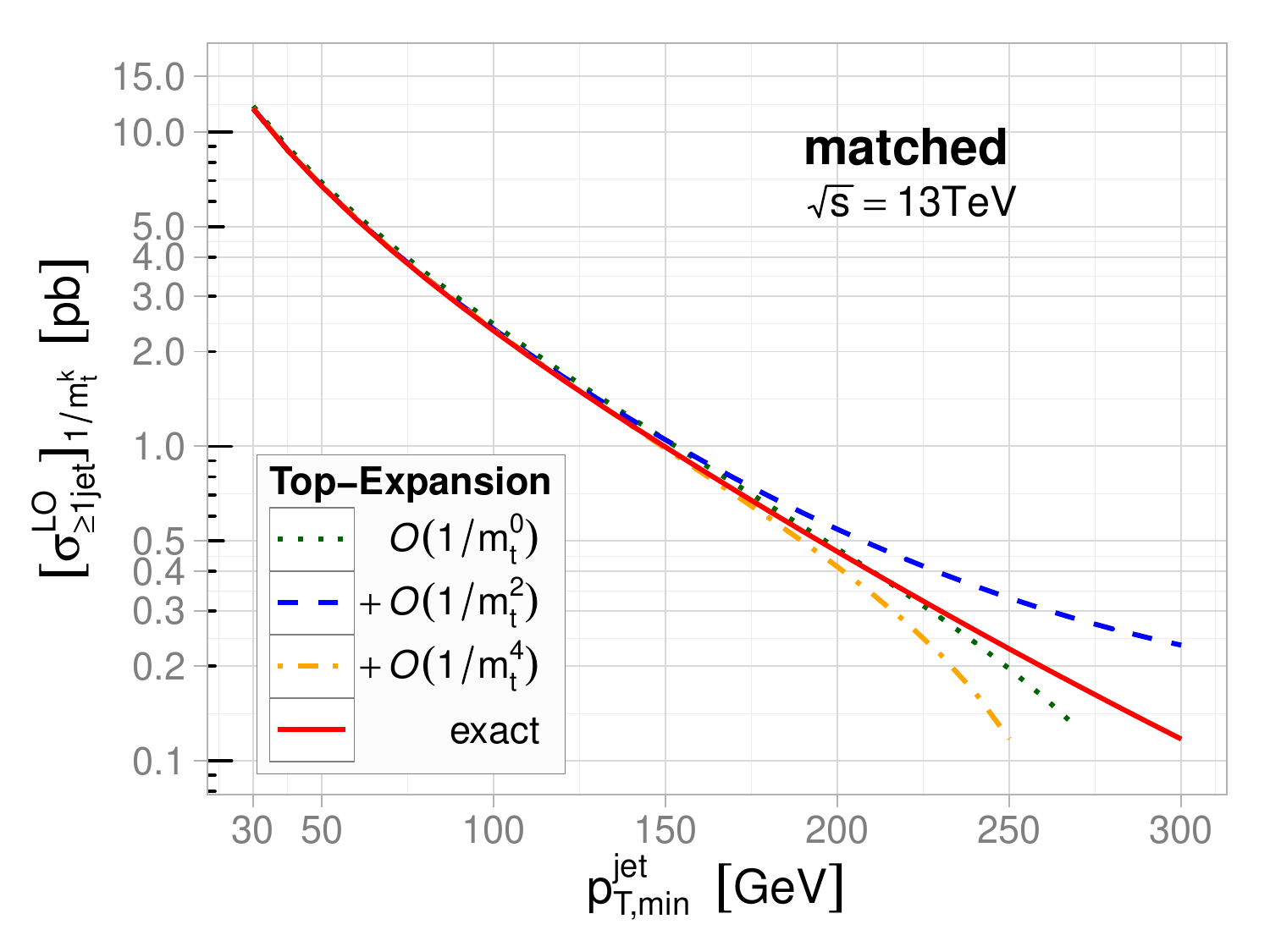}
		\caption{}
        \label{fig:onejetLO_matched}
        \end{subfigure}\\
\begin{subfigure}[b]{0.49\textwidth}
    \includegraphics[width=\textwidth]{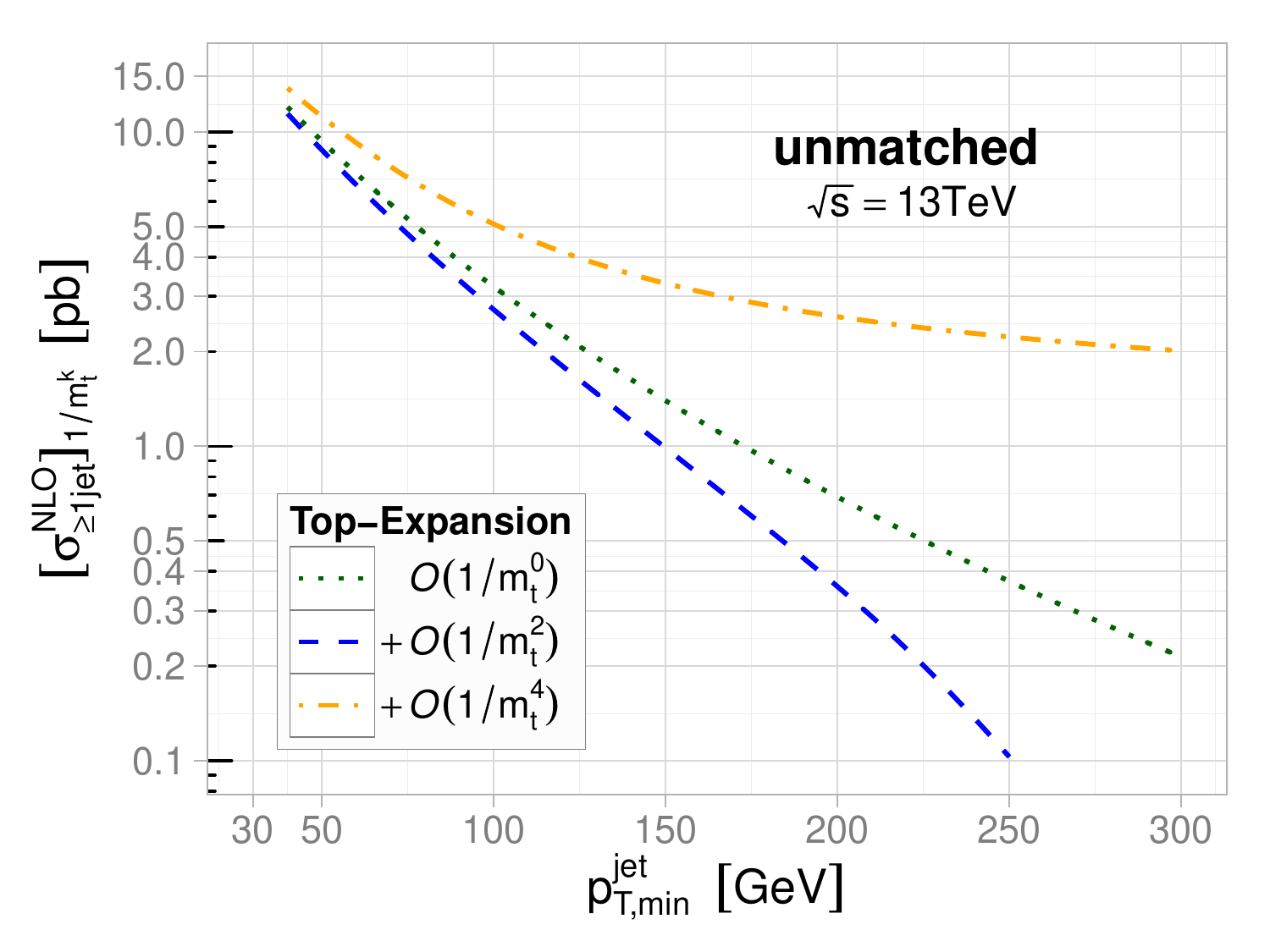}
    \caption{}
    \label{fig:onejet_matched}
\end{subfigure}
\begin{subfigure}[b]{0.49\textwidth}
    \includegraphics[width=\textwidth]{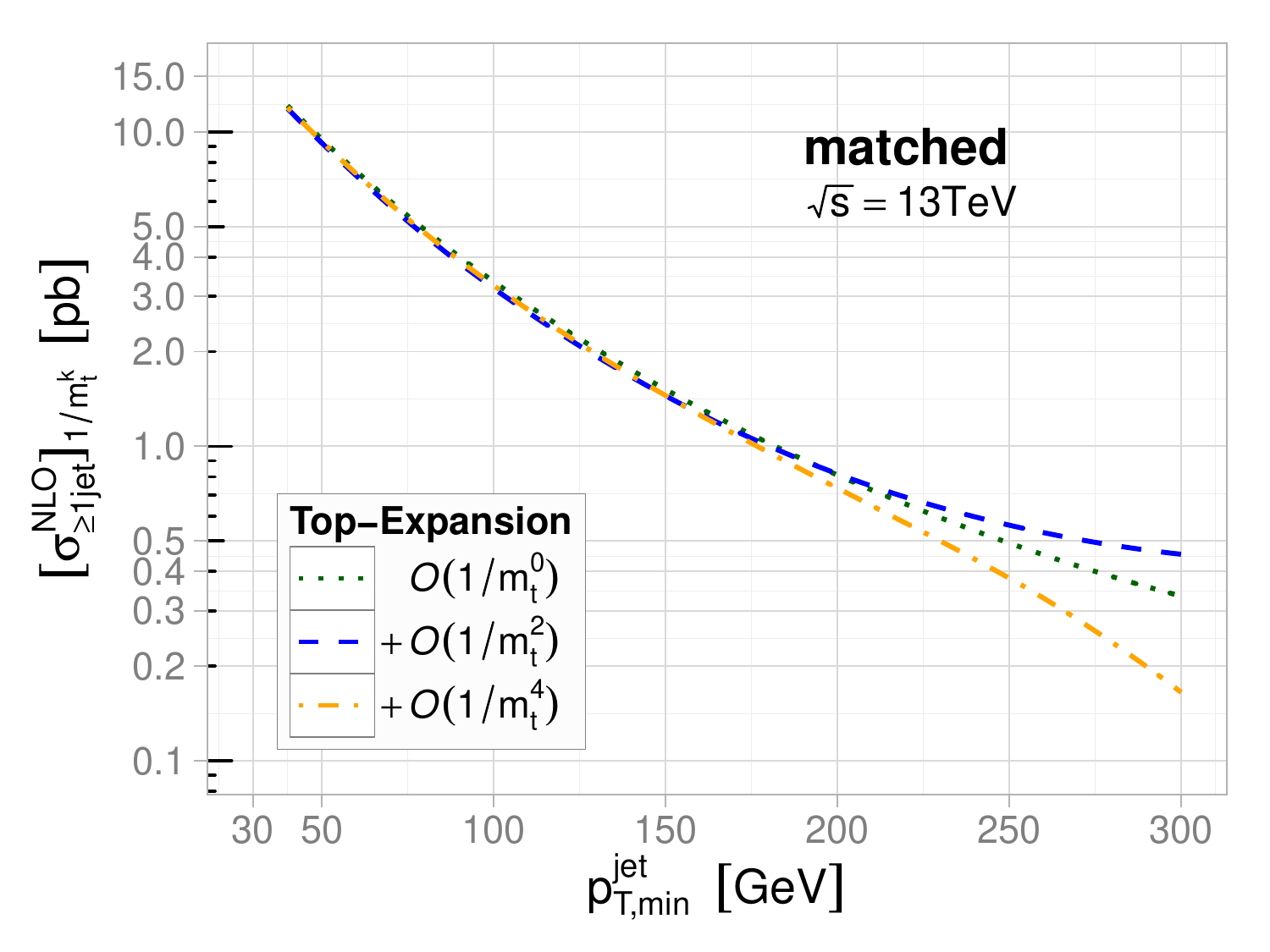}
    \caption{}
    \label{fig:onejet_unmatched}
\end{subfigure}
\begin{center}
\parbox{\textwidth}{%
      \caption[]{\label{fig:HjLO}{Inclusive Higgs+jet rate at (a,b) \lo{}
     and (c,d) \nlo{}
    including terms up to $1/\mtop^k$ as a function of $\ptmin{}$.
    Dotted/dashed/dash-dotted: $k = 0/2/4$, solid: exact. 
    (a,c) unmatched; (b,d) matched according to \eqn{eq:match}.}}}
\end{center}
\end{figure}

It obvious already from \fig{fig:pTH} that in the low-\pt{} region the asymptotic 
expansion is well behaved, while the convergence successively deteriorates 
as \pt{} (and therefore all associated scales) increases. Formally, any event 
with a hardness that exceeds the $2\,\mtop{}$ threshold is outside the validity 
range of the top-mass expansion. Considering the inclusive 
Higgs+jet cross section, 
where at least one hard jet is required,\footnote{In this contribution, 
jets are defined using the anti-$k_T
$ algorithm \cite{Cacciari:2008gp} with jet radius $R = 0.5$ and 
a requirement on the minimal jet transverse momentum of $\ptjet{}>\ptmin$.} the bulk of the 
well-behaved soft region is removed, and the problematic high-scale events
are fully integrated over. For this observable one therefore expects a badly 
converging asymptotic series, as can be seen from \fig{fig:HjLO}\,(a). 
Indeed, none of the approximations agrees with the exact 
\lo{} result (red, solid curve)
and the ordinary $1/\mtop{}$ expansion, including
the heavy-top limit, loses its predictive power.
 
However, one may recover the convergence of the asymptotic expansion, 
by introducing the matched Higgs+jet cross section as follows:
\bal \label{eq:match} \left[\sigma^{\rm matched}_{\ge
1\text{-jet}}\right]_{\mtop^k} \hspace{-0.1cm}\equiv \left[\sigma^{\rm unmatched}_{\ge
1\text{-jet}}\right]_{\mtop^k}  \hspace{-0.1cm}+
\left[\sigma^{\rm matched}_{\text{tot}}\right]_{\mtop^k}  \hspace{-0.1cm}-
\left[\sigma^{\rm unmatched}_{\text{tot}}\right]_{\mtop^k}.
\eal
As can be seen from \fig{fig:HjLO}\,(b), the $1/\mtop^2$ series of the 
matched cross section is nicely convergent and in well agreement with the 
exact \lo{} result at least 
for $\ptmin\lesssim 150$\,GeV.

\eqn{eq:match} stems from the following observation: the total inclusive 
cross section $\sigma_{\text{tot}}$ is integrated over the same problematic 
high-\pt{} region as the Higgs+jet cross section.
As stated before, a matching to the high-energy limit allows to control the 
region $\sqrt{s}>2\,\mtop$ in case of the total cross section \cite{Harlander:2009my} 
(referred to $\sigma^{\rm matched}_{\text{tot}}$). Assuming a similar matching 
for the inclusive Higgs+jet rate 
($\sigma^{\rm matched}_{\ge 1\text{-jet}}$), the difference 
between matched and unmatched cross sections for the total and the inclusive 
Higgs+jet rate should be the same up to a very good precision, 
as long as $\ptmin$ is chosen at values below which 
the asymptotic expansion works well. 
This equality allows us to actually define the matched Higgs+jet cross section 
as done in \eqn{eq:match}. One must be careful, however, to combine the same 
orders in $\alpha_s$ with a consistent set for the \pdf{}s in that equation.

While at \lo{} the exact result is known, the matched 
cross section proves particularly useful for the \nlo{} Higgs+jet rate:
\fig{fig:HjLO}\,(c) shows a similarly bad convergence of the unmatched 
\nlo{} Higgs+jet cross section as observed at \lo{}. With the corresponding 
matching at $\als^4$, shown in \fig{fig:HjLO}\,(d), one 
obtains a nicely behaved asymptotic convergence at least for 
$\ptmin \lesssim 150$\,GeV.
In conclusion, the definition of the matched cross 
section enables a reliable prediction of the \nlo{} Higgs+jet rate
for standard experimental $\ptmin$ cuts.

\subsection{Differential cross sections with resummation}
It is well known that the perturbative ordering in $\als$ breaks down 
in kinematical regions where logarithmically enhanced terms become 
large. One of such regions is $\pt{}\ll Q$, where $Q\sim \mh$ is
the typical hard scale of the Higgs production process. Only a 
resummation of logarithms in $\pt{}/Q$ to all orders in
$\als$ provides a proper theoretical prediction.
Such resummation can be performed analytically or by means of 
a parton shower (\ps{}) approach.

Analytical transverse momentum resummation for the gluon fusion process 
was calculated in the heavy-top approximation
at next-to-next-to-leading logarithmic (\nnll{}) accuracy and consistently 
matched to the \nnlo{} fixed-order cross 
section \cite{Bozzi:2005wk}. This computation is implemented in the 
publicly available programs {\tt HqT}\,\cite{Bozzi:2005wk} and {\tt HRes} 
\cite{deFlorian:2012mx,Grazzini:2013mca}, where the latter is based on a 
Monte Carlo approach that allows for decays of the Higgs boson.

Finite top- and bottom-mass effects on the resummed $\pt$ spectrum were 
discussed in \citere{Bagnaschi:2011tu} for the \powheg{}\,\cite{Nason:2004rx} 
method at \nlo{}\plus{}\ps{}, 
in \citeres{Mantler:2012bj,Grazzini:2013mca,Banfi:2013eda} for analytic
resummation through \nlo{}\plus\nll{} and for \mcatnlo{}\,\cite{Frixione:2002ik} in 
\citere{Mantler:2015vba}. While top-mass effects are 
moderate at small transverse momenta, 
sizable differences were originally observed 
between the \powheg{} prediction for the combined top- and 
bottom-mass effects on the \sm{} \pt{} spectrum and the other two approaches  
\cite{Bagnaschi:2011tu,Mantler:2012bj,Mantler:2015vba}. 

Common to these three approaches (analytic 
resummation, \mcatnlo{}, \powheg) is an effective scale (resummation scale, shower 
scale, $h_{\rm fact}$) that separates the 
soft/collinear from the hard region; referred to as matching scale in the following.
Although the dependence on the matching scales is of higher logarithmic order, 
inadequate values may deteriorate the perturbative convergence due to large 
logarithms, which makes a careful choice necessary. The matching scale is 
 set usually to the characteristic scale of the hard scattering process.
However, if a process involves two or more different mass scales, 
as in the case of the bottom loop in Higgs production ($\mb$ and $\mh$),
this choice is not at all obvious. 
Recently, two proposals were made for the 
algorithmic determination of suitable matching scales separately for 
the top, the bottom and the top-bottom interference contribution
to the cross section \cite{Harlander:2014uea,Bagnaschi:2015qta}.
Indeed, for these matching scale choices the differences
observed in the three resummation/matching approaches  at small \pt{} 
when including 
top- and bottom-mass effects are strongly reduced 
and the predictions become compatible within uncertainties (estimated 
from variations of the central matching scales by a factor of two) \cite{Bagnaschi:2015bop}.
This is true not only in the \sm{}, where the uncertainty induced by 
the bottom loop would not become too severe due to the small bottom Yukawa 
coupling, but also in extended models, where the coupling of the Higgs to 
bottom quarks can be significantly enhanced, and a careful choice of the matching 
scales becomes absolutely crucial \cite{Bagnaschi:2015bop}.

Considering large transverse momenta, \mcatnlo{} and analytic resummation are 
in good agreement due to their transition to the fixed-order prediction at  high \pt{}, 
while \powheg{} develops a considerably larger cross section. The source of this 
enhanced tail is the particular treatment of multiple parton emissions by the \ps{}, 
which acts on all transverse momenta in the ordinary \powheg{} approach.
However, it was shown \cite{Bagnaschi:2015bop} that a simple but powerful modification of the
way \powheg{} is interfaced to the \ps{} leads to a consistent merging with the 
fixed-order prediction in the tail of the distribution and agreement with the other 
two approaches.

\begin{figure}[t]
\begin{subfigure}[b]{0.49\textwidth}
    \includegraphics[width=\textwidth]{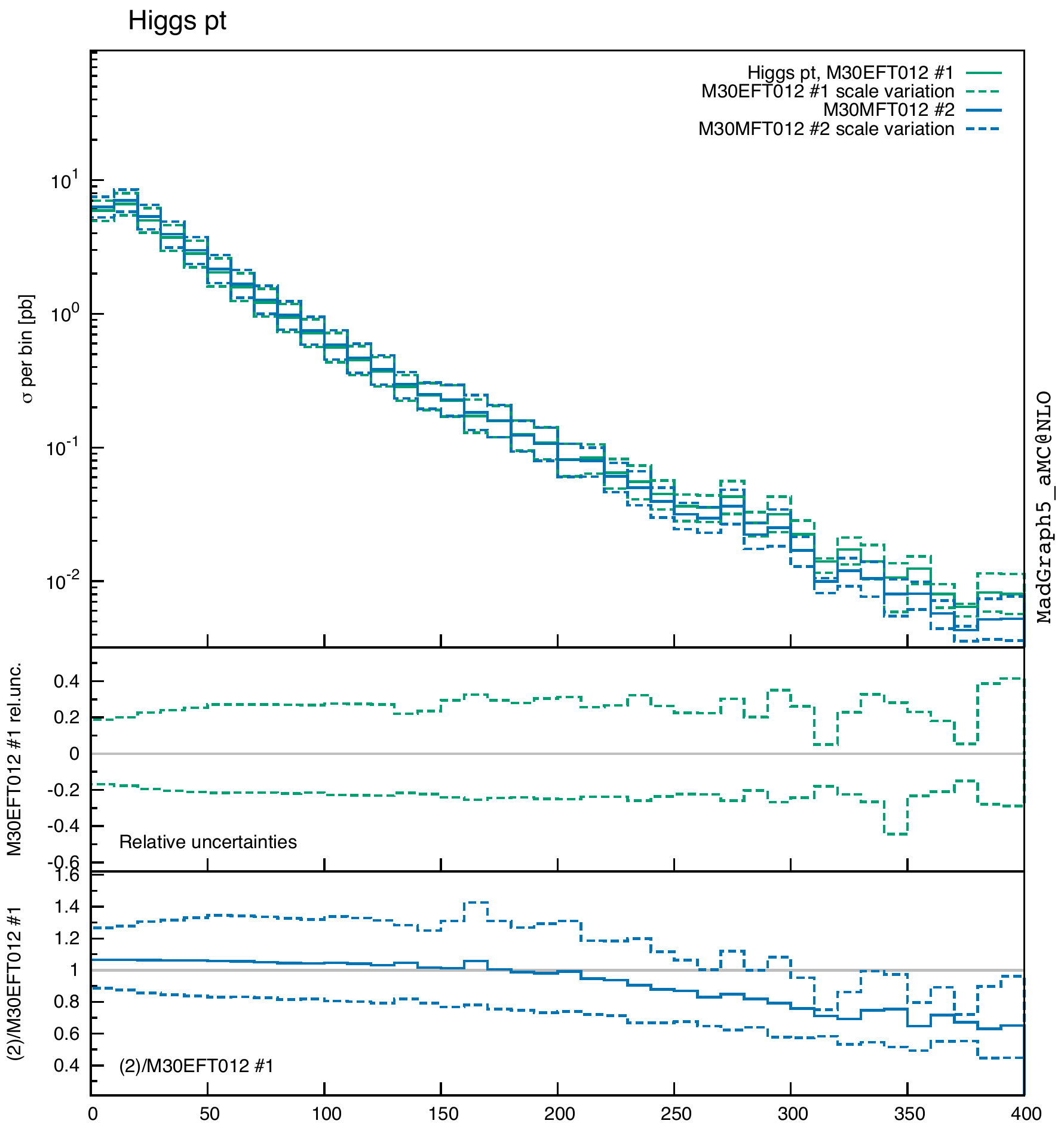}
    \caption{}
    \label{fig:onejet_matched}
\end{subfigure}
\begin{subfigure}[b]{0.49\textwidth}
    \includegraphics[width=\textwidth]{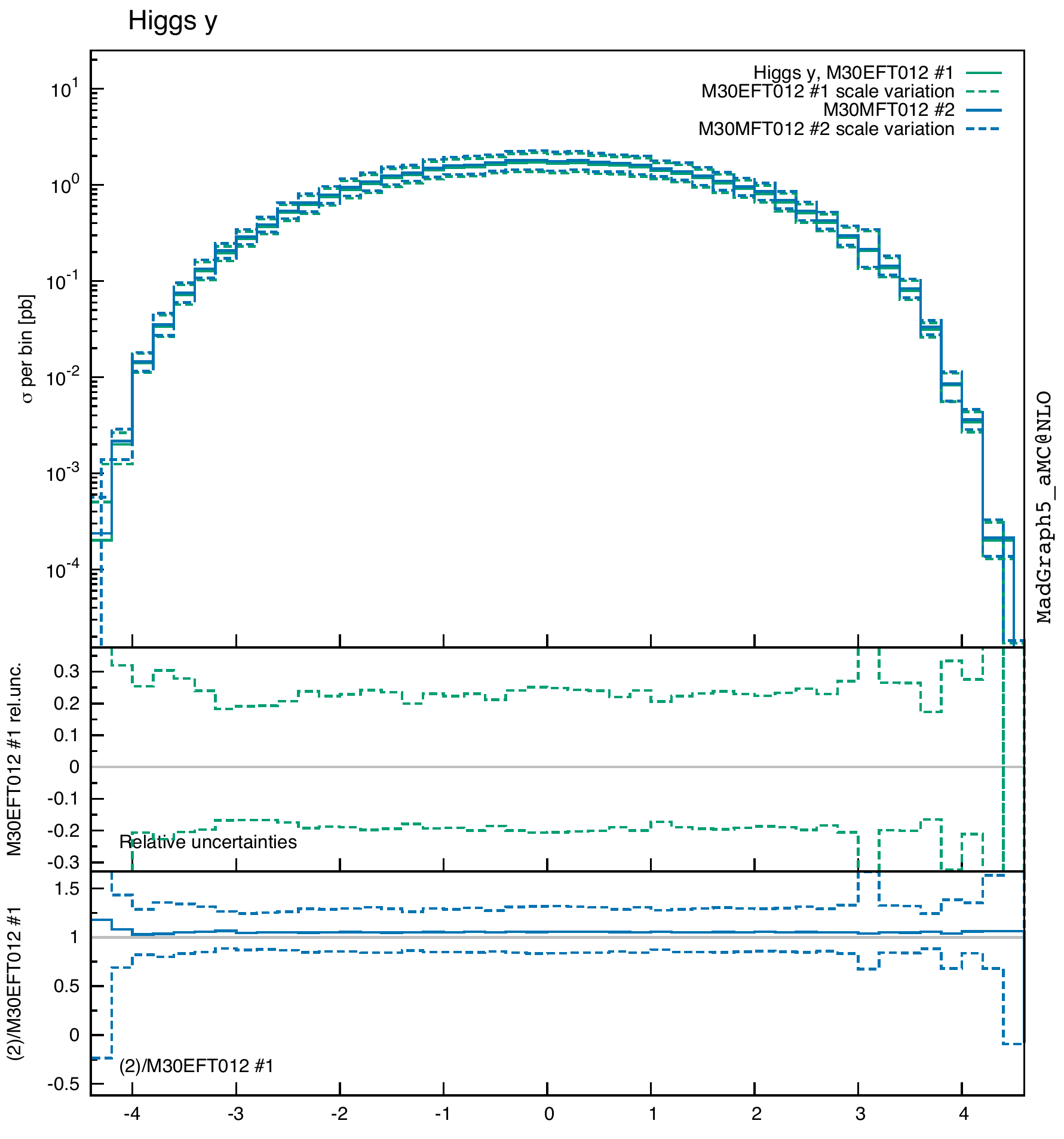}
    \caption{}
    \label{fig:onejet_unmatched}
\end{subfigure}
\begin{center}
\parbox{\textwidth}{%
      \caption[]{\label{fig:merging}{(a) Higgs transverse momentum and 
       (b) rapidity distribution in a merged \nlo\plus\ps{}
      Higgs plus zero, one, two jet computation. Solid light (green) curve: 
      heavy-top limit; solid dark (blue) curve: including mass effects as described in the text.}}} 
\end{center}
\end{figure}

Regarding fully-differential Monte Carlo predictions for hadronic Higgs production
in the \sm{}, a new generation of Monte Carlo tools has been developed in 
the recent past. These computations employ the highest perturbative information 
available in the heavy-top approximation and combine them with finite quark-mass 
effects. They can be divided into two classes: so-called \nnlo{}\plus\ps{} approaches 
\cite{Hamilton:2013fea,Hoche:2014dla,Hamilton:2015nsa}, 
which merge Higgs plus zero and one jets at \nlo{}\plus{}\ps{}, while including \nnlo{} 
corrections to the inclusive Higgs cross section and rapidity distribution; and 
\nlo{}\plus\ps{} merged computations of Higgs plus zero, one and two jets 
\cite{Frederix:2012ps,FFVW,Hoeche:2014lxa,Buschmann:2014sia}.

The currently most complete computation \cite{FFVW} 
in terms of heavy-quark mass effects 
 employing all available exact matrix elements up to Higgs plus 
three jets, has been implemented by means of FxFx merging 
\cite{Frederix:2012ps} in the {\tt MadGraph5\_aMC@NLO} 
framework \cite{Alwall:2014hca}. The only approximated matrix 
elements are the two-loop virtuals for Higgs plus one and two jets, which are 
computed in the heavy-top limit and improved by reweighting them with 
the born amplitude in the full theory. \fig{fig:merging} compares the transverse 
momentum distribution (left panel) and rapidity distribution (right panel) of the 
Higgs boson in the heavy-top approximation (solid light, green curve) to the prediction 
including top-mass effects (solid dark, blue curve). From the ratio between the blue and the 
green curve in the second inset it is obvious, that top-mass effects become 
particularly relevant at large transverse momentum scales and have a significant 
impact on the \pt{} shape, while for the rapidity distribution they essentially only affect the normalization. Bottom-mass effects can be 
added solely for Higgs plus zero jets 
at \nlo{}\plus{}\ps{}, since the heavy-quark approximation provides no adequate 
description in that case.

\bibliographystyle{polonica-mod}
\bibliography{M_Wiesemann}

\end{document}